\documentclass[aps,pra,twocolumn,footinbib]{revtex4-1}
\usepackage{amsmath,amsfonts,amssymb}
\usepackage{graphicx,float,calc}
\usepackage{color,bm}
\usepackage[colorlinks,urlcolor=blue,citecolor=blue,linkcolor=blue]{hyperref}
\usepackage{amsmath}
\usepackage{amsfonts}
\usepackage{amssymb}
\usepackage{graphicx}
\usepackage{mathrsfs}
\usepackage[sort&compress]{natbib}

\begin{document}

\title{Multicriticality and quantum fluctuation in generalized Dicke model}
\author{Youjiang Xu, Diego Fallas Padilla, Han Pu}

\begin{abstract}
We consider an important generalization of the Dicke model in which
multi-level atoms, instead of two-level atoms as in conventional Dicke
model, interact with a single photonic mode. We explore the phase diagram of
a broad class of atom-photon coupling schemes and show that, under this
generalization, the Dicke model can become multicritical. For a subclass of
experimentally realizable schemes, multicritical conditions of arbitrary
order can be expressed analytically in compact forms. We also calculate the
atom-photon entanglement entropy for both critical and non-critical cases.
We find that the order of the criticality strongly affects the critical
entanglement entropy: higher order yields stronger entanglement. Our work
provides deep insight into quantum phase transitions and multicriticality.
\end{abstract}

\maketitle

\affiliation{Department of Physics and Astronomy, and Rice Center for
Quantum Materials, Rice University, Houston, Texas 77251-1892, USA}

The Dicke model \cite{PhysRev.93.99} is one of the most iconic models in
quantum optics and quantum many-body physics. It describes an ensemble of
two-level atoms interacting with a single photonic mode. When the
atom-photon coupling strength exceeds a threshold, the system enters the
superradiance phase via a second-order phase transition where the $Z_{2}$
symmetry of the model is spontaneously broken, and the photonic mode is
macroscopically populated. The Dicke model and the superradiant quantum
phase transition (SQPT) have been realized in various experimental settings,
including quantum gases of neutral atoms \cite%
{PhysRevA.75.013804,PhysRevLett.104.130401,baumann2010dicke,PhysRevA.97.043858}%
, trapped ion system \cite{PhysRevLett.121.040503}, super-conducting circuit
\cite{lamata2017digital,mezzacapo2014digital,langford2017experimentally},
and solid state systems \cite{Li794}.

Real atoms, of course, possess complicated level structure. Even if we
restrict ourselves to the ground state manifold, a typical atom often
features more than two levels. This motivates our current work to
investigate an important generalization of the Dicke model where the
two-level atoms are replaced by multi-level atoms. As we shall show, this
generalized Dicke model also supports the SQPT in the thermodynamic limit
\footnote{%
SQPT is also present in the so-called classical oscillator limit, where the
frequency of the bosonic mode tends to zero \cite{PhysRevA.85.043821}. In
the classical oscillator limit, the phase transition shares the same
mean-field description as that in the thermodynamic limit, but the
underlying quantum fluctuation is different. We will discuss the classical
oscillator limit of the generalized Dicke model in a separate paper.}, but
the order of the phase transition can now be controlled and multicritical
points may emerge. We will provide a detailed study on the condition of the
emergence of the multicritical points of arbitrary order, and show that
higher order multicritical points are associated with a higher degree of
atom-photon entanglement.

Our multicritical Dicke model describes $N$ $l$-level atoms coupled with a
single photonic mode of frequency $\omega $. The Hamiltonian can be written
as ($\hbar=1$):%
\begin{equation}
H=\omega a^{\dag }a+\frac{g\left( a+a^{\dag }\right) }{2\sqrt{N}}%
\sum_{k=1}^{N}d^{\left( k\right) }+\epsilon \sum_{k=1}^{N}\, h^{\left(
k\right) },  \label{H}
\end{equation}%
where $a$ is the photon annihilation operator, dimensionless single-atom
Hamiltonian $h$ and dipole operator $d$ act on the $l$ inner states of
atoms, $g$ and $\epsilon $ set the energy scales of the atom-photon
interaction and the internal energy of the atoms, respectively.

Multicritical points are special points on the critical manifold.
Multicriticality is defined by deduction, i.e., an $n^{\mathrm{th}}$-order
critical manifold is the boundary of the $\left( n-1\right)^{\mathrm{th}}$%
-order critical manifold, and the ordinary critical points are defined to be
2$^{\mathrm{nd}}$ order \footnote{%
Exactly speaking, an $n^{\mathrm{th}}$-order critical manifold is the
intersection of multiple $\left( n-1\right)^{\mathrm{th}}$-order critical
manifolds, some of which, though, may not show up in the physically
accessible phase space, e.g., excluded by the $Z_{2}$ symmetry in the model
we study.}\cite{PhysRevB.8.346}. For example, a tricritical point can appear
where a discontinuous phase transition boundary and a 2$^{\mathrm{nd}}$%
-order critical line smoothly intersect \cite%
{PhysRevLett.24.715,PhysRevB.8.346,chaikin1995principles}. Multicritical
points often belong to a universality class different from that of the
ordinary critical points \cite{PhysRevLett.28.675,henkel2013conformal}, so
they may provide new insight into quantum phase transitions. Especially,
because quantum criticality is characterized by increasing quantum
fluctuation, we want to study how the quantum fluctuation is affected by
multicriticality. However, systems that support high-order critical points
are rarely found, for the reason that more than a few parameters need to be
fine-tuned, and it is usually challenging to pinpoint the multicritical
manifold in the phase space because the dimension of the manifold decreases
when the order of criticality increases. Usually, the dimension of an $n^{%
\mathrm{th}}$-order critical manifold is $\left( n-1\right) $ less than the
number of tunable parameters \cite{PhysRevB.8.346}. It turns out that
Hamiltonian~(\ref{H}) serves as a great platform to investigate
multicriticality because: (1) It provides plenty of tunable parameters that
are experimentally realistic; and (2) it is possible to derive exact
analytical expressions of multicritical conditions, the equations that
determines the multicritical manifolds.


We will first discuss the multicritical Dicke model within the mean-field
framework. The mean-field Hamiltonian $H_{\mathrm{MF}}$ is obtained by
replacing the bosonic operator $a$ with a real number $\epsilon \sqrt{N}\phi
/g$:%
\begin{equation}
H_{\mathrm{MF}}/\epsilon =\kappa \phi ^{2}+\phi d+h\text{ ,}  \label{MF}
\end{equation}%
where $\kappa :=\omega \epsilon /g^{2}$. We denote the eigenstates of $H_{%
\mathrm{MF}}$ as $\left\vert k\right\rangle $'s with eigenvalues $\epsilon
_{k} $ $(k=1,2,\dots ,l)$. We assume that $|1 \rangle$ is the non-degenerate
ground state of $H_{\mathrm{MF}}$. The mean-field ground state energy is
obtained by minimizing $\epsilon _{1}$ with respect to $\phi $, and the $%
\phi $ that minimizes $\epsilon _{1}$ is recognized as the order parameter.
The SQPT occurs in the thermodynamic limit $N\rightarrow \infty $ when the
order parameter becomes non-zero.

The $Z_2$ symmetry of Hamiltonian (\ref{H}), which is spontaneously broken
by the SQPT, manifests itself as $H$ is invariant under the transformation $%
a\rightarrow -a$, $d\rightarrow -d$, $h\rightarrow h$. Given this symmetry,
the mean-field ground state energy can be written as a Taylor series in
terms of $\phi ^{2}$: $\epsilon _{1}=\sum_{k=0}^{\infty }c_{k}\phi ^{2k}$.
An ordinary critical point is met when $c_{1}=0$ and $c_{2}>0$. The boundary
of this manifold satisfies the condition $c_{1}=c_{2}=0$, which specifies
the tricritical manifold. Generally, the $n^{\mathrm{th}}$-order critical
manifold satisfies the condition $c_{1}=c_{2}=\cdots =c_{n-1}=0$. These
multicritical conditions can be expressed as equations on $d$ and $h$ using
perturbation theory. Treating $h$ as the unperturbed Hamiltonian and $\phi d$
as the perturbation, we obtain $c_{k}$ by carrying out the perturbation
expansion to $(2k)^{\mathrm{th}}$ order. For example, the 2$^{\mathrm{nd}}$%
-order perturbation expansion recasts $c_{1}=0$ as
\begin{equation}
\sum_{k=2}^{l}\frac{\left\vert d_{1k}\right\vert ^{2}}{h_{kk}}=\kappa \text{%
, }  \label{ordinary}
\end{equation}%
where the matrix elements are taken respect to the eigenvectors of $h$ that
satisfies $h_{kk}>h_{11}=0$ for $k=2,3,\dots,l$. The 4th-order perturbation
recasts $c_{2}=0$ as:
\begin{equation}
\sum_{k_{1}k_{2}k_{3}=2}^{l}\frac{%
d_{1k_{1}}d_{k_{1}k_{2}}d_{k_{2}k_{3}}d_{k_{3}1}}{%
h_{k_{1}k_{1}}h_{k_{2}k_{2}}h_{k_{3}k_{3}}} =\sum_{k_{1}k_{2}=2}^{l}\frac{
\left\vert d_{1k_{1}}\right\vert ^{2}\left\vert d_{1k_{2}}\right\vert ^{2}}{%
h_{k_{1}k_{1}}^{2}h_{k_{2}k_{2}}}\text{.}  \label{c2}
\end{equation}
and so on so forth.

In general, the existence of the $n^{\mathrm{th}}$-order critical points
requires at least $\left( n-1\right) $ tunable parameters. For the
multicritical Dicke model considered here, the number of internal atomic
levels $l$ and the $Z_{2}$ symmetry put constraints on the number of tunable
parameters. The $Z_{2}$ symmetry requires the presence of a parity operator $%
P$ which makes $PdP=-d$ and $PhP=h$. Suppose the number of $\pm 1$ in the
eigenvalues of $P$ is $\frac{l\pm \delta }{2}$. If we represent $d$ and $h$
as matrices using a set of common eigenvectors of $h$ and $P$ as basis, then
$h$ is diagonal and contains $l$ tunable parameters, which are just the
eigenvalues of $h$; whereas $d$ must be in the form $d=%
\begin{pmatrix}
0 & M \\
M^{\dag } & 0%
\end{pmatrix}%
$ where $M$ is an arbitrary $\frac{ l-\delta}{2} \times \frac{ l+\delta}{2}$
matrix. Now we have $\frac{l^{2}-\delta ^{2}}{2}+l$ parameters, besides, we
have to fix the $\left( l-1\right) $ relative phases between the common
eigenvectors, also the physics would not change if we rescale $H$ or shift
the zero point energy. In the end, we have at most $G=\left( l^{2}-\delta
^{2}\right) /2-1$ tunable parameters. For example, for two-level atoms with $%
l=2$ as in the conventional Dicke model, we have $\delta =0$ and $G=1$,
which means no multicriticality. In order to find multicriticality, we must
have at least $l=3$. We note that a tricritical point is identified in a
spin-1 Bose gas subjected to spin-orbit coupling \cite{campbell2016magnetic}%
. This system can be recast into the form of the generalized Dicke model
with $l=3$ in the classical oscillator limit. In our previous work \cite%
{PhysRevLett.122.193201}, by introducing a staggered magnetic field to the
two-level atoms, we show that this modified Dicke model exhibits
tricriticality. This model can be regarded as a special case of the
generalized Dicke model with $l=4 $.

Although the procedure of finding the multicritical condition of any order
is straightforward under the perturbation approach outlined above, Eqs.~(\ref%
{ordinary}) and (\ref{c2}) indicate that these equations quickly become very
complicated as the order increases. Even numerical solutions to these
equations may become impractical. However, we will show now that, for a
subclass of the multicritical Dicke models, we can write down the
multicritical conditions to arbitrary order in compact analytic forms. For
this subclass, still under the representation where $h$ is diagonal, only
the super- and sub-diagonal elements of the $d$ matrix are non-vanishing,
i.e., $d_{ij}= 0$ if $|i-j| \neq 1$. As a result, $H_{\mathrm{MF}}$ takes a
tridiagonal form, and hence we call this subclass as the T-class. For a
T-class Hamiltonian, the $n^{\mathrm{th}}$-order critical condition is given
by the simple form
\begin{equation}
\left\vert d_{k,k-1}\right\vert ^{2}=\kappa h_{kk}\,,\;\;\;\text{ for }2\leq
k\leq n\text{.}  \label{critical condition}
\end{equation}%
To prove it, we denote, for a given $l$, the determinant of $H_{\mathrm{MF}}$
as $\zeta _{l}$, which can be expressed as a Taylor series of $\phi ^{2}$.
If $c_{1}=c_{2}=\cdots =c_{n-1}=0$, then $\zeta _{l}\propto \phi ^{2n}$. We
shall now prove that $\zeta _{l}\propto \phi ^{2n}$ as long as Eq.~(\ref%
{critical condition}) holds. To this end, we write down the recurrence
relation for $\zeta _{k}$ by exploiting the tridiagonal form of $H_{\mathrm{%
MF}}$:
\begin{equation*}
\zeta _{k}=\left( h_{kk}+\kappa \phi ^{2}\right) \zeta _{k-1}-\phi
^{2}\left\vert d_{k,k-1}\right\vert ^{2}\zeta _{k-2}\text{ . }
\end{equation*}%
Under the condition of Eq.~(\ref{critical condition}), we have $\zeta
_{2}=\kappa ^{2}\phi ^{4}$, $\zeta _{3}=\kappa ^{3}\phi ^{6}$. By deduction,
it is easy to prove $\zeta _{k}=\kappa ^{k}\phi ^{2k}$ for $k=2,3,\dots ,n$.
As a result, $\zeta _{l}$ must be proportional to $\phi ^{2n}$ for $n \leq l$%
, which finishes the proof.

The experimental scheme for realizing a T-class Hamiltonian has been
proposed in \cite{PhysRevLett.119.213601} and realized in \cite{Zhiqiang:17}%
. Using the $F=2$ hyperfine ground state of $^{85}$Rb with cavity-assisted
Raman transitions, it is possible to realize up to 5th-order criticality
following the critical conditions Eq.~(\ref{critical condition}). With one
pair of Raman lasers as proposed in \cite{PhysRevLett.119.213601}, the
relative strength between $d_{n,n-1}$'s are fixed as $\left(
d_{12},d_{23},d_{34},d_{45}\right) =\left( \sqrt{2},\sqrt{3},\sqrt{3},\sqrt{2%
}\right) $ 
, which will always be used in our numerical
studies. With this $d$ matrix, the multicritical conditions are met by
tuning $h_{kk}$'s, which represent the bare energies of the atomic internal
states and which can be tuned with external magnetic fields via the Zeeman
shift or external microwave fields via the AC-Stark shift \cite%
{PhysRevA.73.041602,PhysRevA.79.043631,PhysRevLett.107.195306,PhysRevA.5.968,PhysRevA.75.053606,PhysRevA.79.023406,PhysRevLett.111.185305}%
. We set $\kappa = 1$ in our numerical calculation, then the 5th-order
critical point is located at $\left(h_{22},h_{33},h_{44},h_{55}\right)
=\left( 2,3,3,2\right)$. 

\begin{figure}[tbh]
\centering
\includegraphics[width=0.48\textwidth]{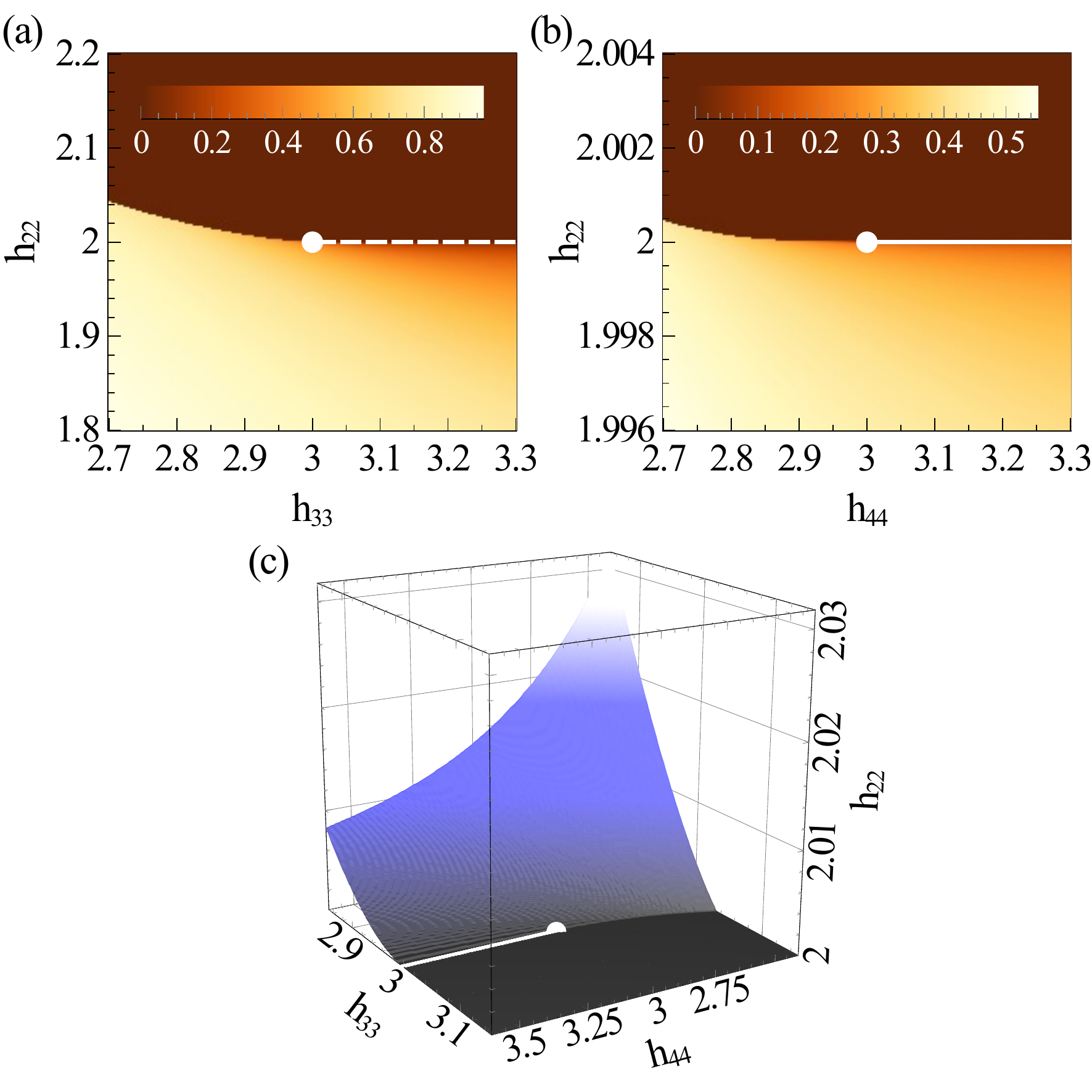}
\caption{(Color online) (a)(b) The mean-field phase diagram of a 4-level
T-class Hamiltonian around the 4th-order critical point. The colorbar
represents the order parameter $\protect\phi$. The dashed line in (a) is a
ordinary critical line while the solid line in (b) is a tricritical line.
(c) The boundary between the normal phase and the superradiant phase, in
which $h_{22}$ is plotted as a function of $h_{33}$ and $h_{44}$. The $%
h_{22}=2$ plane is the critical manifold, with a tricritical line marked by
a solid line. The curved surface is the 1st-order phase transition boundary.
In all panels, we use a white dot to mark the 4th-order critical point.}
\label{phaseDiagram}
\end{figure}

In Fig.~\ref{phaseDiagram}, we plot the phase diagrams of the T-class
Hamiltonians that are experimentally available. In Fig.~\ref{phaseDiagram}%
(a)(b), we show the change of the order parameter with respect to $h$, which
is tuned around a tetracritical (i.e., 4$^{\mathrm{th}}$-order) point. This
is done by setting $h_{55}$ to be very large, hence we have effectively a
4-level atomic system. The tetracritical point is located at $%
(h_{22},h_{33},h_{44})=(2,3,3)$ and is marked by the white dot in the
graphs. In Fig.~\ref{phaseDiagram}(a), we fix $h_{44}=3$ and vary $h_{22}$
and $h_{33}$. The darker region represents the normal phase and the lighter
region the superradiance region. The ordinary $2^{\mathrm{nd}}$-order
critical line is marked by the white dashed line which is a straight line
with $h_{22}=2$ and $h_{33}>3$. The phase boundary to the left of the
tetracritical point is of $1^{\mathrm{st}}$-order. In Fig.~\ref{phaseDiagram}%
(b), we fix $h_{33}=3$ and vary $h_{22}$ and $h_{44}$. Here the straight
solid line with $h_{22}=2$ and $h_{44}>3$ is a tricritical line, which joins
the $1^{\mathrm{st}}$-order boundary at the tetracritical point. In Fig.~\ref%
{phaseDiagram}(c), we show the boundary surface between the normal (above
the surface) and the superradiant phase (below the surface) in the full
three dimensional parameter space. This boundary surface contains two parts:
a flat part at plane representing the $2^{\mathrm{nd}}$-order critical
manifold and a curved part representing the $1^{\mathrm{st}}$-order surface.
The tricritical line and the tetracritical point are marked by the white
solid line and the white dot, respectively.

Having discussed the mean-field phase diagram, we now turn our attention to
the quantum fluctuation and the entanglement properties of the model. By
shifting the bosonic operator in Eq.~(\ref{H}) by the mean-field order
parameter, $b:=a-\epsilon \sqrt{N}\phi /g$, Eq.~(\ref{H}) is rewritten as%
\begin{equation}
H=\omega b^{\dag }b+\frac{g\left( b^{\dag }+b\right) }{2\sqrt{N}}%
\sum_{k=1}^{N}D^{\left( k\right) }+\sum_{k=1}^{N}H_{\mathrm{MF}}^{\left(
k\right) } \,,  \label{shifted H}
\end{equation}%
where the shifted dipole operator is $D:=d+2\kappa \phi $.

Without loss of generality, the atomic states can be expressed in terms of
an orthonormal basis consisting of completely symmetrized Fock states $%
\left\vert \mathbf{\chi }\right\rangle $, where $\mathbf{\chi }$ is a vector
whose component $\chi _{i}$ denote the number of atoms occupying $\left\vert
i\right\rangle $, the $i^{\mathrm{th}}$ eigenstate of $H_{\mathrm{MF}}$.
Consequently, the last term in Eq.~(\ref{shifted H}) is diagonal in this
basis, $\left\langle \mathbf{\chi |}\sum_{k=1}^{N}H_{\mathrm{MF}}^{\left(
k\right) }\mathbf{|\chi }\right\rangle =\sum_{i=1}^{l}\epsilon _{i}\chi _{i}$%
. The non-zero matrix elements of $\sum_{k=1}^{N}D^{\left( k\right) }$ under
the basis $\left\vert \mathbf{\chi }\right\rangle $'s are
\begin{align}
\left\langle \mathbf{\chi }\right\vert \sum_{k=1}^{N}D^{\left( k\right)
}\left\vert \mathbf{\chi }\right\rangle & =\sum_{k=1}^{N}\chi _{k}D_{k,k}%
\text{ ,} \\
\left\langle \mathbf{\chi }^{i,j}\right\vert \sum_{k=1}^{N}D^{\left(
k\right) }\left\vert \mathbf{\chi }\right\rangle &=\sqrt{\chi _{j}\left(
\chi _{i}+1\right) }D_{i,j}\text{ ,}
\end{align}%
where $\left\vert \mathbf{\chi }^{i,j}\right\rangle $ is the very state
containing one more atom in $\left\vert i\right\rangle $ and one less atom
in $\left\vert j\right\rangle $ than $\left\vert \mathbf{\chi }\right\rangle
$. As long as we are interested in the low-energy states, we can assume that
most atoms occupy the mean-field ground state $\left\vert 1\right\rangle$,
i.e., $\chi _{1}\sim N$ and $\chi _{k}=o\left( N\right) $ for $k=2,3,\dots
,l $. We can then express $H=H_{\mathrm{eff}}+N\epsilon _{1}+o\left(
1\right) $ when $N\rightarrow \infty $, and the low-energy effective
Hamiltonian $H_{\mathrm{eff}}$ is quadratic in $b$ and new bosonic operators
$b_{2},b_{3},\dots ,b_{l}$:
\begin{equation}
H_{\mathrm{eff}}=\omega b^{\dag }b+\sum_{i=2}^{l} \left[\omega
_{i}b_{i}^{\dag }b_{i}+\frac{g}{2}\left\vert D_{1,i}\right\vert \left(
b+b^{\dag }\right) \left( b_{i}+b_{i}^{\dag }\right) \right]\text{, }
\label{Heff}
\end{equation}%
where $b_{i}$ is defined by $b_{i}\left\vert \mathbf{\chi }%
^{1,i}\right\rangle =\sqrt{\chi _{i}}\left\vert \mathbf{\chi }\right\rangle $%
, and $\omega _{i}=\epsilon _{i}-\epsilon _{1}$. In deriving Eq.~(\ref{Heff}%
), we have used $D_{1,1}=0$, which results from the steadiness of the
mean-field energy $\partial _{\phi }\epsilon _{1}=0$. Because the leading
term in the asymptotic series of $H$ is the mean-field energy $N\epsilon
_{1} $, it serves as a confirmation that the mean-field theory determines
the exact phase diagram as long as the asymptotic expansion is valid. The
validity of the expansion can be verified self-consistently, i.e., it is
valid as long as $\sum_{i=2}^l\left\langle b_{i}^{\dag }b_{i}\right\rangle
\ll N$, where the expectation value is taken in respect to the low-energy
states.

To find the ground state of $H_{\mathrm{eff}}$, the effective Hamiltonian
can be transformed into a Hamiltonian describing an $l$-dimensional harmonic
oscillator,%
\begin{equation}
H_{\mathrm{eff}}=\frac{1}{2}\sum_{j,k=1}^{l}\left( P_{j}\delta
_{jk}P_{k}+X_{j}\Omega _{jk}^{2}X_{k}\right) -\frac{1}{2}\sum_{k=1}^{l}%
\omega _{k}\text{, }
\end{equation}%
where $P_{k}=\sqrt{\frac{\omega _{k}}{2}}\left( b_{k}-b_{k}^{\dag }\right) $
and $X_{k}=\frac{1}{\sqrt{2\omega _{k}}}\left( b_{k}+b_{k}^{\dag }\right) $
are canonical momentum and position operators as linear combinations of $%
b_{k}$ and $b_{k}^{\dag }$. Here we have denoted $b_{1}\equiv b$ and $\omega
_{1}\equiv \omega $. The squared eigenfrequencies of the harmonic oscillator
are given by the eigenvalues of the matrix $\Omega ^{2}$. The non-zero
matrix elements of $\Omega ^{2}$ are given by%
\begin{align*}
\Omega _{kk}^{2}& =\omega _{k}^{2}\,,\text{ for }k=1,2,\dots ,l, \\
\Omega _{1k}^{2}& =\left\vert D_{1,k}\right\vert \sqrt{\omega \omega _{k}}\,,%
\text{ for }k>1\text{.}
\end{align*}%
The eigenvalues of $\Omega ^{2}$ are determined by the characteristic
polynomial
\begin{equation*}
p\left( \lambda ^{2}\right) =\left( 1-\sum_{k=2}^{l}\frac{\left\vert
D_{1,k}\right\vert ^{2}\omega \omega _{k}}{\left( \omega ^{2}-\lambda
^{2}\right) \left( \omega _{k}^{2}-\lambda ^{2}\right) }\right)
\prod_{k=1}^{l}\left( \omega _{k}^{2}-\lambda ^{2}\right) \text{, }
\end{equation*}%
If $\omega _{j-1}\neq \omega _{j}=\omega _{j+1}=\cdots =\omega _{j+f}<\omega
_{j+f+1}$, then $\lambda ^{2}=\omega _{j}^{2}$ must be an $f$-fold
eigenvalue. The corresponding eigenstates are trivial in the sense that they
are dark states which do not couple to the light mode. In studying the SQPT,
we are not interested in these dark states so we remove the degeneracy by
requiring $\omega _{j}<\omega _{j+1}$ for $j>1$. Now the characteristic
equation is equivalent to
\begin{equation*}
q\left( \lambda ^{2}\right) :=\omega ^{2}-\lambda ^{2}+\omega \sum_{k=2}^{l}%
\frac{\left\vert D_{1,k}\right\vert ^{2}\omega _{k}}{\lambda ^{2}-\omega
_{k}^{2}}=0\text{ ,}
\end{equation*}%
from which we notice the distribution of the solutions: $\lambda _{1}^{2}\in
\left( -\infty ,\min \left( \omega ^{2},\omega _{2}^{2}\right) \right) $, $%
\lambda _{2}^{2}\in \left( \omega _{2}^{2},\omega _{3}^{2}\right) ,\cdots $,
$\lambda _{l}^{2}\in \left( \omega _{l}^{2},+\infty \right) $. We note that
only $\lambda _{1}$ can be zero as long as the mean-field ground state is
non-degenerate. The asymptotic expansion is valid when $\lambda _{1}^{2}>0$,
otherwise the fluctuation blows up. Consequently, the equation $q\left(
0\right) =0$ gives the critical condition, which coincides with the one
derived through the mean-field theory in Eq.~(\ref{ordinary}).

The ground state wave function of $H_{\mathrm{eff}}$ is an $l$-dimension
Gaussian function, whose atom-photon entanglement entropy can be calculated
straightforwardly:
\begin{align}
S& =\frac{\gamma }{e^{\gamma }-1}-\ln \left( 1-e^{-\gamma }\right) , \\
\gamma & =\cosh ^{-1}\left( \frac{\Omega _{11}M_{11}+\det \Omega }{\Omega
_{11}M_{11}-\det \Omega }\right) ,
\end{align}%
where $M_{11}$ is the $\left( 1,1\right) $-minor of the matrix $\Omega $.
Here $S$ is the von Neumann entropy of the reduced density matrix for either
the atomic or the photonic modes. Because the critical condition can also be
expressed as $\det \Omega =0$ or $\gamma =0$, the entropy diverges at the
critical points. Near the critical points, $\gamma $ is small, so
approximately we have $S=1-\ln \gamma $, which indicates a logarithm
divergence approaching the critical point. The entropy near the critical
points is closely related to the fluctuation in the light mode, which can be
read from the following equation%
\begin{equation}
\left\langle \left( b^{\dag }+b\right) ^{2}\right\rangle =\frac{\omega M_{11}%
}{\det \Omega }.
\end{equation}%
In addition, the entropy is closely related to the first excitation gap $%
\lambda _{1}=\det \Omega /\prod_{i=2}^{l}\lambda _{k}$.

\begin{figure}[htb]
\centering
\includegraphics[width=0.48\textwidth]{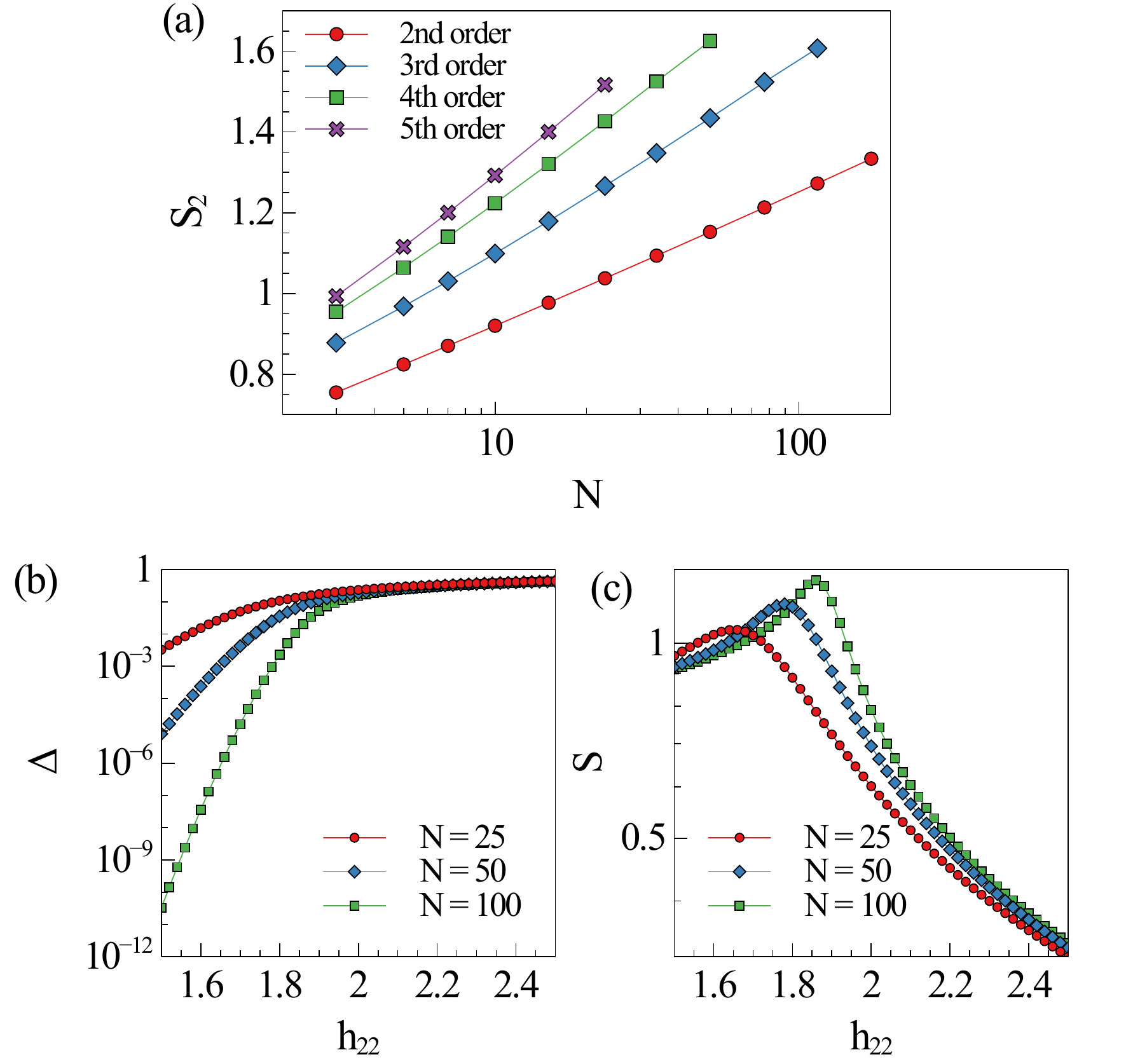}
\caption{(Color online) (a) Critical atom-photon entanglement entropy $S_{%
\mathrm{cri}}$ plotted against the atom number $N$ for T-class Hamiltonians
with different orders of criticality. (b) The gap $\Delta$ between the
ground state and the first excited state, and (c) the ground state entropy $S
$, of the two-level Dicke model with $d_{12}=\protect\sqrt{2}$ for different
atom number $N$. $\Delta$ exponentially decreases when $h_{22}$ moves deep
into the superradiant phase. }
\label{finiteNEntropy}
\end{figure}

Finally, we examine the entanglement entropy at the critical points where
the asymptotic expansion possibly fails. We numerically calculate the finite-%
$N$ critical entropy for the T-class Hamiltonians. In the thermodynamic
limit, the ground state is non-degenerate in the normal phase while doubly
degenerate in the superradiant phase, and the excitation gap closes at the
phase boundary and remains closed in the superradiance region. For finite $N$%
, however, the gap does not close, but rather decreases exponentially when
we move deeper into the superradiant region. Therefore, the phase boundaries
and the critical manifolds in a finite system cannot be determined
unambiguously by the gap, as demonstrated in Fig.~\ref{finiteNEntropy}(b).
Instead, we locate the critical manifold for finite $N$ by maximizing the
atom-photon entanglement entropy $S$ by varying $h_{22}$, as shown in Fig. ~\ref{finiteNEntropy}(c). The maximized
entropy is identified as the critical entropy $S_{\text{cri}}$ for finite $N$%
.

\begin{table}[tbp]
\caption{Asymptotic behavior of the critical entropy $S_{\mathrm{cri}}\sim s_{0}+s_{1}\ln N$%
.}
\label{finiteNEntropytable}%
\begin{ruledtabular}
\begin{tabular}{ccccc}
Order of Criticality & 2 & 3 & 4 & 5\\
$s_{0}$ & 0.593(2) & 0.642(7) & 0.682(7) & 0.704(8)\\
$s_{1}$ & 0.1428(7) & 0.207(2) & 0.238(3) & 0.257(3)\\
\end{tabular}
\end{ruledtabular}
\end{table}

In Fig.~\ref{finiteNEntropy}(a), we plot $S_{\mathrm{cri}}$ against $N$ at
critical points with different order of criticality for the T-class
Hamiltonians. Asymptotically, we have $S_{\mathrm{cri}}\sim s_{0}+s_{1}\ln N$%
. The fitting parameters $s_{0}$ and $s_{1}$ are shown in Table.~\ref%
{finiteNEntropytable}, from which we see that higher-order criticality is
associated with a larger degree of entanglement. Different orders of
criticality are achieved by setting $\left( h_{33},h_{44},h_{55}\right)
=\left( 3,3,2\right) $ then tuning some $h_{kk}$ to infinity. For example,
to achieve the tricriticality, we set $h_{44}=h_{55}=\infty $.

\emph{Conclusion ---} We replace the two-level atoms in conventional Dicke
model with $l$-level atoms and study the superradiance phase transition in
the modified model. The increased number of tuning parameters for $l>2$
leads to the emergence of multicriticality whose order can be controlled.
The phase diagram and the multicritical conditions can be obtained from the
mean-field theory. For subclass of the multicritical Dicke models, which can
be readily realized experimentally, we show that the multicritical
conditions of arbitrary order can be expressed analytically in compact
forms. The non-critical atom-photon entanglement entropy of the
multicritical Dicke models in the thermodynamic limit can be calculated
analytically through an asymptotic expansion of the Hamiltonian. The entropy
diverges logarithmically when approaching the critical point. The entropy at
the critical points for finite number of atoms are calculated numerically.
We found that the critical entropy increases when the order of criticality
increases. Our work provides deep insights into the physics of quantum phase
transition and multicritical points, whose realization is typically very
challenging in other contexts.

We acknowledge support from the NSF and the Welch Foundation (Grant No.
C-1669).
%

\end{document}